\documentclass[letterpaper,dvips]{eptcs}

\title{Higher-order Rewriting for Executable Compiler Specifications}
\author{
  Kristoffer H. Rose
  \institute{IBM Thomas J.~Watson Research Center\\
    P.O. Box 704, Yorktown Heights, NY 10598, USA}\\
  \email{krisrose@us.ibm.com}
}
\def\titlerunning{Higher-order Rewriting for Compilers}
\def\authorrunning{K.H. Rose}

\usepackage{rcs}
\RCS$Id: on-compilers.tex,v 1.5 2011/02/13 23:46:21 krisrose Exp $
\RCS$Revision: 1.5 $
\RCS$Date: 2011/02/13 23:46:21 $

\usepackage[utf8]{inputenc}
\usepackage{xcolor,amsmath,amsfonts,amsthm,stmaryrd,verbatim,breakurl}
\usepackage[all]{xy}
\theoremstyle{definition}

\let\lt=< \let\gt=>

\newcommand{\eg}{\textit{e.g.}}
\newcommand{\etc}{\textit{etc.}}
\newcommand{\meta}[2][]{\ensuremath{\textit{#2}_{#1}}}

\DeclareUnicodeCharacter{00AC}{\ensuremath{\lnot}}
\DeclareUnicodeCharacter{03BB}{\ensuremath{\lambda}}
\DeclareUnicodeCharacter{2020}{\ensuremath{\dagger}}
\DeclareUnicodeCharacter{2021}{\ensuremath{\ddagger}}
\DeclareUnicodeCharacter{2192}{\ensuremath{\rightarrow}}
\DeclareUnicodeCharacter{21D2}{\ensuremath{\Rightarrow}}
\DeclareUnicodeCharacter{2200}{\ensuremath{\forall}}
\DeclareUnicodeCharacter{2203}{\ensuremath{\exists}}
\DeclareUnicodeCharacter{2261}{\ensuremath{\equiv}}
\DeclareUnicodeCharacter{22A2}{\ensuremath{\vdash}}
\DeclareUnicodeCharacter{22A4}{\ensuremath{\top}}
\DeclareUnicodeCharacter{22A5}{\ensuremath{\bot}}
\DeclareUnicodeCharacter{22A8}{\ensuremath{\vDash}}
\DeclareUnicodeCharacter{27E6}{\ensuremath{\llbracket}}
\DeclareUnicodeCharacter{27E7}{\ensuremath{\rrbracket}}
\DeclareUnicodeCharacter{27E8}{\ensuremath{\langle}}
\DeclareUnicodeCharacter{27E9}{\ensuremath{\rangle}}

\SelectTips{eu}{11}
\definecolor{dkmagenta}{rgb}{.4,0,.4}
\definecolor{dkgreen}{rgb}{0.,.5,.0}
\definecolor{dkblue}{rgb}{0.,.0,.5}
\hypersetup{colorlinks,linkcolor=dkblue,urlcolor=dkmagenta,citecolor=dkgreen}

\begin{document}
\maketitle

\begin{abstract}\noindent
  In this paper we outline how a simple compiler can be completely specified using higher order
  rewriting in all stages: parsing, analysis/optimization, and code emission, specifically using the
  \emph{crsx.sf.net} system for a small declarative language called ``X'' inspired by XQuery (for
  which we are building a production quality compiler in the same way).
\end{abstract}


\section{Introduction}

A compiler typically consists of a parser generating an abstract syntax tree (AST) for some source
language (SL), a ``normalization'' to a canonical form in an intermediate language (IL), some
rewrites inserting analysis results into and performing simplifications of the IL, and finally code
emission to the target language (TL).
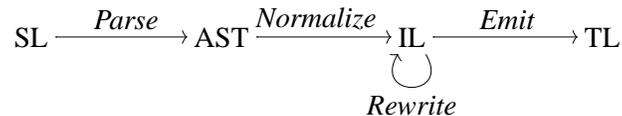
\begin{displaymath}
  \xygraph{ !~*{+\txt} !{0;(2,0):0}
    []{SL} :^{\hbox{\it Parse}}[r] {AST}="AST"
    :^{\hbox{\it Normalize}}[r] {IL}="IL"
    :@(dr,dl)^{\hbox{\it Rewrite}} "IL"
    :^{\hbox{\it Emit}} [r] {TL}
  }
\end{displaymath}
Each arrow in the diagram can be understood as a rewriting:
\begin{enumerate}

\item parsing to an AST is a rewriting from the string of characters in the input file to a term
  representing the source program, usually formalized and implemented using some variation of
  context free grammars~\cite{Knuth:1968:MST};

\item normalization of the AST into the IL involves rewrite rules to eliminate ``syntactic sugar''
  and other redundant aspects of the source language;

\item rewriting of the IL involves adding annotations, simplifications, and sometimes using parts of
  the program itself like rewrite rules (for example for inlining defined functions); often some
  rewrites depend on the result of other rewrites (like an optimization depending on an analysis);
  finally,

\item code emission is usually a direct expansion of the ``finished'' IL program into sequences (or
  templates) of instructions that are directly executable by a computer.

\end{enumerate}
We'll show how each of these steps is specified using the CRSX system~\cite{Rose:hor2007,CRSX}, an
implementation of a variation of Combinatory Reduction Systems~\cite{Klop+:tcs1993}.  The actual
samples we'll present below are mere toys, of course, but they do illustrate the ideas in a manner
that is consistent with a production compiler that we are building for XQuery~\cite{w3c:xquery}.

We first summarize the CRSX system notation, including the extensions, in Section~\ref{crsx}, before
we introduce the parser specification in Section~\ref{pg} followed by the normalizer rules in
Section~\ref{N}.  Section~\ref{R} then explains a few simple sample rewrites, and Section~\ref{E}
presents code emission rules.  Finally, we conclude and discuss some related work in
Section~\ref{Wrap}.


\section{CRSX Summary}\label{crsx}

Our setting is \emph{Combinatory Reduction Systems}~\cite{Klop+:tcs1993} as realized by the ``CRSX''
system~\cite{CRSX}.  Here we briefly summarize the used notation and where it differs from reference
CRS.

Terms are constructed from the basic grammar
\begin{align}
  \tag{Terms} t &::= v \mid \{e\}C[s,\dots,s] \mid \{e\}M[t,\dots,t] \\
  \tag{Scope} s &::= \vec{v} . t \mid t \\
  \tag{Environment} e &::= M \mid e;\,v:t \mid e;\,C:t
\end{align}
where variables, $v$, are written with a lower case letter (including composite units like
\verb|v"$x"|), meta-variables, $M$, must include a hash mark (\verb|#|) in the name, and all other
units (including literal constants) are constructors, $C$.

Term formation is as shown, where constructions $\{e\}C[s,\dots,s]$ are non-standard in two ways:
\begin{itemize}

\item Each subterm of a construction is a \emph{scope}, which may include a vector of distinct
  variable ``binders'' ($\vec{v}$ denotes $v_1 v_2 \dots v_n$ for $n\gt0$), which can then occur as
  variables inside the scope (with the usual caveat that the innermost possible scope is used for
  each particular variable name; this is the only location where the formalism accepts abstraction).

\item Each construction has an associated \emph{environment} component, which is a collection of
  mappings from constructors and variables to terms (in addition to permitting meta-variables for
  pattern matching against environments).

\end{itemize}
Meta-applications $\{e\}M[t,\dots,t]$ are used in rewrite rules of the form
\begin{displaymath}
  \meta{name}[\meta{options}] : \meta{pattern} → \meta{contraction}
\end{displaymath}
with the following extended version of the CRS conventions:
\begin{itemize}

\item The \meta{name} becomes the name of the rule; it can be replaced with ``\verb|-|'' to use a
  default name.

\item The \meta{options} is a comma-separated list of instructions to relax the requirement that all
  used meta-variables occur exactly once on each side of the rule, that all variables are explicitly
  scoped, and that all pattern meta-applications permit all in-scope variables (to avoid accidental
  $\eta$-style rules).

\item The \meta{pattern} is a term that must be a construction wherein contained meta-applications
  are applied exclusively to distinct bound variables.  The pattern defines what the rule will
  \emph{match}: specifically the rule will match any subterm where the top constructor matches
  including have the same number of parameters and binders on the parameter scopes, matching all
  required environment members, and matching the shape of each parameter term recursively with the
  addition that pattern meta-applications match any corresponding parameter term provided only the
  included bound variables occur in the matched term (as usual for CRS; we give examples later).
  The mapping from the meta-variables with the parameter bound variables to the real term and its
  bound variables is called a \emph{valuation}; CRSX extends valuations to also map \emph{whole
    environment meta-variables} and \emph{free variables} to parts of the matched term.

\item The $→$ is the Unicode U2192 character.

\item The \meta{contraction} explains what the matched subterm should be replaced with by the
  rewrite step.  Constructions stand for themselves.  Meta-applications stand for copies of what the
  meta-variable matched where in turn the matched bound variables are \emph{substituted} by the
  corresponding arguments provided in the contraction meta-application, in usual CRS fashion.
  Variables bound in the contraction just stand for themselves but free variables either stand for
  occurrences of the variable they matched or, as a special feature can be declared ``fresh,'' which
  means a new globally unique fresh variable is created~\cite{Rose:1996}.  Environments in the
  contraction can reference matched environment meta-variables extended with additional bindings.

\end{itemize}
Finally, the CRSX parser permits the following abbreviations borrowed from $λ$ calculus and
programming languages:
\begin{itemize}

\item Parenthesis are allowed around every term, so $\verb"("t\verb")"$ is the same as~$t$;

\item $c~\vec{v}.t$ abbreviates $c[v_1.c[v_2.\,\cdots\,c[v_n.t]\cdots]]$ (think $λxyz.t$);

\item $t_1t_2$ abbreviates \verb"@["$t_1$\verb","$t_2$\verb"]" and is left recursive so $t_1t_2t_3$
  is the same as $(t_1t_2)t_3$;

\item $t_1\verb";"t_2$ abbreviates \verb"$Cons"\verb"["$t_1$\verb","$t_2$\verb"]" and is right
  recursive with the addition that omitted segments correspond to \verb"$Nil", so
  \verb"("$t_1$\verb";"$t_2;$\verb")" corresponds to the term
  \verb"$Cons["$t_1$\verb",$Cons["$t_2$\verb",$Nil]]"; 
  and

\item empty brackets \verb"[]" can be omitted.

\end{itemize}


\begin{figure*}[p]\small
\begin{verbatim}
// Grammar for X (simple XQuery-like language).
class net.sf.crsx.samples.x.X : <P>, <E>, <S>, <Q>

meta[<E>] ::= "#<PRODUCTION_NAME>" i?, "[", "]" . // Meta-applications over AST.

skip ::= " " | "\r" | "\n" | "\t" .               // White space.

<P> ::= {program} <E> .                           // Program.

<E> ::= <S>:#S (","_$ ⟦#S⟧ <E> | ⟦#S⟧) .           // Expression.


<S> ::= "(" (<E> | {empty}) ")"                   // Simple expression.
     |  "element"_$ <N> "{" <E> "}"
     |  {query} <Q>
     |  "if"_$ <S> "then" <S> "else" <S>
     |  {call} <N> "(" (<E> | {empty}) ")"
     |  v_?
     |  {literal} <L>
     .

<Q> ::= "for"_$ v_x "in" <S> <Q>[x]              // Query.
     |  "let"_$ v_x ":=" <S> <Q>[x]
     |  "where"_$ <S> <Q>
     |  "return"_$ <S>
     .

token v ::= "$" n .                              // Variable tokens.

<N> ::= n_$ .                                    // Names.
token n ::= [A-Za-z_] [A-Za-z0-9_-]* .

<L> ::= l_$ .                                    // Literals.
token l ::= i | s .
token i ::= [0-9]+ .
token s ::= "'" (¬[\'] | "''")* "'" .
\end{verbatim}
\caption{\textit{x.pg}---parsing X to AST.}\label{fig:pg}
\end{figure*}
\begin{figure*}[t]\small
\begin{verbatim}
// for $x in child(doc()) for $y in child(doc()) where eq($x,$y) return plus($x,$y)
"program"[
 "query"[
  "for"[
   "call"["child", "call"["doc", "empty"]],
   v"$x" .
    "for"[
     "call"["child", "call"["doc", "empty"]],
     v"$y" .
      "where"[
       "call"["eq", ","[v"$x", v"$y"]],
       "return"["call"["plus", ","[v"$x", v"$y"]]]]]]]]
\end{verbatim}
\caption{Example parse from X program to AST.}\label{fig:AST-sample}
\end{figure*}

\section{Parser}\label{pg}

The first component of our X compiler is the parsing from X syntax to the AST, which are terms in a
higher-order abstract syntax representation~\cite{PfenningElliot:sn1988} of X.  Thus the parser has
to be instructed for every production in the language how the AST subterm for that production must
look, including what binders should be introduced and how they can occur. Figure~\ref{fig:pg} shows
the actual file used to achieve this with the CRSX system's PG parser generator.  (Note that like
all files used by the CRSX system, the parser generator file is a Unicode text file which permits us
to use special characters.)

The grammar itself is specified as follows:
\begin{itemize}

\item \texttt{//} introduces comments.

\item The first line declares the external ``class'' name we'll use for the parser as well as the
  default and other externally visible non-terminals that the parser can be explicitly requested to
  parse.

\item The rest of the file consists of units that start with a name or some special keyword and end
  with a period.

\item The unit starting with \texttt{meta} gives the special notation used for meta-variables when
  writing rules involving parsed expressions; we'll return to this in the following section and here
  just remark that we use a notation for meta-variables inside parsed text which is a subset of the
  CRSX meta-variable notation, and the unit starting with \texttt{skip} declares the white space
  convention.

\item In general, non-terminals are written in angle brackets, like \texttt{<P>}, terminals (or
  defined tokens) are written as simple identifiers, like \texttt{v}, and literal tokens are written
  as strings like \texttt{","}.

\item Units starting with a non-terminal name are the proper productions.  In productions,
  non-terminals and terminals stand for themselves, we use parenthesis \texttt{()} for grouping, and
  vertical bar \texttt{|} for choice---all else is annotations, explained below.  So the first two
  productions could have been written as
\begin{verbatim}
   <P> ::= <E> .
   <E> ::= <S> ("," <E> |) .
\end{verbatim}
  if we were not interested in generating an AST term.

\item Units starting with \texttt{token} give the regular expression for the defined token.  We use
  conventional regular expression syntax with character classes written in \texttt{[]}s (negated by
  a preceding \texttt{¬} and including ranges), choice with \texttt{|}, optionality and repetition
  with \texttt{?+*}, and literal characters as strings.

\end{itemize}
The purpose of the parsing, however, is to build an AST for the parsed X program.  This is achieved
by the annotations in the productions.
\begin{itemize}

\item The default behaviour is that tokens are (parsed but) ignored and non-terminals are parsed and
  submitted as subterms to the current context.

\item When a production includes a name in braces, like \verb|{program}| in the \texttt{<P>}
  production, this specifies that the production generates an AST term with the tag \texttt{program}
  with all following subtrees as children (specifically up to the end of the current choice).

\item When a token is followed by \texttt{\_\$} then this specifies that an AST term using the token
  as the tag with all following subtrees as children (up to the end of the surrounding choice).  So
  in the \texttt{<N>} production, the \texttt{n} token is directly used as the tag (with no children
  since there are no following parsed non-terminals).

\item When a generated subterm is followed by a colon (\texttt{:}) and a meta-variable name starting
  with a hash \texttt{\#}, then this means that the subtree generated from the non-terminal is not
  echoed to the context but stored with that name for later use in an inserted term in double
  brackets ⟦\dots⟧.  So, for example, we can read the \texttt{<E>} production as follows:
  \begin{enumerate}
  \item Parse the \texttt{<S>} subterm and remember it as \texttt{\#S} instead of including it in
    the context.
  \item If the next token is a comma then the result is a term rooted by a comma-tag and with two
    subtems: the one stored as \texttt{\#S} generated by \texttt{⟦\#S⟧} and the one generated by the
    following \texttt{<E>}.
  \item Otherwise, the result is just what was stored as \texttt{\#S} generated by (the second)
    \texttt{⟦\#S⟧} without any additional tag.
  \end{enumerate}
  (The fact that a tag can be omitted is a powerful feature that permits us to confuse the
  \texttt{<E>} and \texttt{<S>} non-terminals in the normalization rules, as we shall see.)

\item An annotation of \texttt{\_x}, where \texttt{x} can be any lower case variable name
  identifier, promotes the token value to a \emph{scoped identifier definition} and makes
  \texttt{[x]} after a single non-terminal in the same production indicate that the scope of
  \texttt{x} is that non-terminal.  So, for example, in the first choice of the \texttt{<Q>}
  production, the \texttt{v} token is used as a variable name which is scoped in the \texttt{<Q>}
  subterm.

\item Finally, \texttt{\_?} after a token indicates that the token must be an \emph{occurrence} of a
  bound variable.

\end{itemize}
In summary, the parser specification looks like many other abstract syntax tree generation
notations, such as MetaPRL~\cite{HickeyNogin:hosc2006} or ANTLR Tree Grammars~\cite{Parr:2008:tree},
except for the additional direct support for higher order abstract syntax by explicitly specifying
the scoping and a pleasantly compact way to generate terms where tokens are used directly as
constructors, which reduces the size of large parsers considerably.

Figure~\ref{fig:AST-sample} shows a sample AST printed by the CRSX engine for the term shown in the
comment.
The generated tags are quoted because they would otherwise be mistaken for other CRSX syntax;
similarly, actual CRSX variables that do not start with a lower case letter are written as
\texttt{v"\$x"}, \etc, which allows us to retain the original X names in the AST.  Notice how the
AST term binds two variables, one for each \texttt{"for"} construct, following the CRSX constraint
that binders are only permitted on construction subterms.


\begin{figure*}[p]\small
\begin{verbatim}
// N: Normalization scheme: compile from X AST to nested relational algebra IL.
N[(
$CheckGrammar['net.sf.crsx.samples.x.X'] ; // we need to parse X fragments

// Program.
N[%P⟦ #E ⟧] → Algebraic[Dep[id.N[#E, id]]] ;

// Expressions: N[expression, input-tuple] rewrites to operator.

-[Free[id]] : N[%E⟦ ( #S , #E )     ⟧, id] → Concat[N[#S, id], N[#E, id]] ;
-[Free[id]] : N[%S⟦ ()              ⟧, id] → Empty ;
-[Free[id]] : N[%S⟦ #L              ⟧, id] → Literal[#L] ;
-[Free[id]] : N[%S⟦ element #N {#E} ⟧, id] → Element[Literal[#N], N[#E, id]] ;
-[Free[id]] : N[%S⟦ #N(#E)          ⟧, id] → Call[#N, N[#E, id]] ;

-[Free[id]] : N[%S⟦ if #S then #S1 else #S2 ⟧, id]
 → Conditional[N[#S, id], N[#S1, id], N[#S2, id]] ;

-[Free[f,id]] : N[f, id] → Extract[id, f] ;

// Queries.
-[Free[id]] : N[%S⟦ #Q ⟧, id] → NQ[#Q, id, t.t] ;

// NQ[query source, input-variable, t.prefix-operator[t]]

-[Free[id],Fresh[f]] :
NQ[%Q⟦ for $v in #S #Q[$v] ⟧, id, t.#op[t]]
 → NQ[#Q[f], id, id3.MapConcat[Dep[id2.
                       Map[Dep[id1.Tuple[ACons[f id1, ANil]]], N[#S, id2]]], #op[id3]]] ;

-[Free[id],Fresh[f]] :
NQ[%Q⟦ let $v := #S #Q[$v] ⟧, id, t.#op[t]]
 → NQ[#Q[f], id, id2.MapConcat[Dep[id1.Tuple[(f N[#S, id1];)]], #op[id2]]] ;

-[Free[id]] :
NQ[%Q⟦ where #S #Q ⟧, id, t.#op[t]]
 → NQ[#Q, id, id2.Select[Dep[id1.N[#S, id1]], #op[id2]]] ;

-[Free[id]] :
NQ[%Q⟦ return #S ⟧, id, t.#op[t]]
 → Map[Dep[id1.N[#S, id1]], #op[id]] ;

)]
\end{verbatim}
\caption{\textit{N.crs}---normalizing X terms to nested-relational algebra.}\label{fig:N}
\end{figure*}
\begin{figure*}[t]\footnotesize
\begin{verbatim}
Algebraic[
 (Dep id .
  Map[
   (Dep id1 . Call["plus", Concat[Extract[id1, v"$x"], Extract[id1, v"$y"]]]),
   Select[
    (Dep id1_1 . Call["eq", Concat[Extract[id1_1, v"$x"], Extract[id1_1, v"$y"]]]),
    MapConcat[
     (Dep id2 .
      Map[(Dep id1_2 . Tuple[ACons[(v"$y" id1_2), ANil]]), Call["child", Call["doc", Empty]]]),
     MapConcat[
      (Dep id2_1 .
       Map[(Dep id1_3 . Tuple[ACons[(v"$x" id1_3), ANil]]), Call["child", Call["doc", Empty]]]),
      id]]]])]
\end{verbatim}
\caption{Normalized version of sample query.}
\label{fig:N-out}
\end{figure*}

\section{Normalization}\label{N}

Our sample intermediate language is a variant of nested-relational
algebra~\cite{RothEtal:1988:tods,CluetMoerkotte:1993:dblp} modified to make the binders of dependent
operators explicit so we can exploit the higher-order rewriting capabilities, \eg, we write the map
operator as
\begin{displaymath}
  \mathtt{Map[Dep[id.}p_2\mathtt{],}\,p_1\mathtt{]}
\end{displaymath}
with an explicit \texttt{Dep} dependency abstraction to scope the ``context tuple'' (usually denoted
by a context sensitive symbol like \texttt{ID} in relational algebra).

The actual normalization rules are shown in Figure~\ref{fig:N}, and exercise most of the features of
CRSX:
\begin{itemize}

\item We first check that we have the grammar from Section~\ref{pg} loaded.  The grammar enables two
  notations:
  \begin{enumerate}

  \item In CRSX syntax, \texttt{\%P⟦\dots⟧} denotes \emph{inline parsing} of the \dots\ text using
    the \texttt{<P>} production of some grammar (that must have been loaded in advance).

  \item Inside parsed text, \texttt{\#P} denotes \emph{any} subterm where a \texttt{<P>} subterm is
    allowed; for disambiguation, such subterms further permit a numeric marker like \texttt{\#P2}.
    (This is what the \texttt{meta} declaration in Figure~\ref{fig:pg} is for.)

  \end{enumerate}
  The first rule then expresses that a \texttt{<P>}-program containing an \texttt{<E>} subterm (they
  all do) rewrites to the shown \texttt{Algebraic}-term, where the \texttt{N}-subterm is the one
  representing the compilation scheme that will lead to the entire AST being normalized recursively.

\item Notice that the right hand side of the first rule introduces a binder: \texttt{id} is bound in
  the invocation of \texttt{N}.  In all the rules for \texttt{N} we shall explicitly refer to this
  variable, however, in those cases it will (locally) be a \emph{free} variable where we do not know
  the binder.

  Thus all the following rules include the \emph{option} \texttt{Free[id]} to indicate that the
  pattern can use \texttt{id} to match a free variable.  (This is otherwise not permitted as it is
  likely to be the result of mistyping.)

  Matching of free variables in this way is inherently problematic for confluence, because it breaks
  the confluence of developments: if the variable is substituted by something then the rule no
  longer applies!  Thus we need an assurance that \emph{variables that are matched against and
    substitued are disjoint}.  For the present system this is ensured by the AST data structures
  being pure input data in the sense that no rule produces an AST construction, and no AST data is
  allowed to escape from the \verb|N| wrapper.  (The other way to ensure non-substitution is to
  create globally fresh free variables since only bound variables can be substituted.)

\item The next block of rules defines all the easy cases of normalization of sequences, literals,
  element creation, function calls, conditional, and finally field extraction, which does not
  involve any X syntax because all fields are converted to free field tag variables, as we shall
  see.

\item Finally, queries are translated backwards~\cite{GhelliEtal:2007:dbpl} using an ``operator
  accumulator'' third argument with the \texttt{NQ} helper compilation scheme.  The first two rules
  of the \texttt{NQ} scheme involve replacing a bound variable with a globally fresh one, which is
  achieved by the use of higher-order matching and rewriting:
  \begin{enumerate}
  \item the pattern of the rules includes the fragment \texttt{\#Q[\$v]}, which establishes that the
    \texttt{<Q>} subterm should be matched with ``tracking'' of all occurrences of the variable
    bound by the \texttt{for} or \texttt{let} construct, respectively (the notation used here is
    determined by the \texttt{meta} declaration in the parser description file);
  \item the rules include the option \texttt{Fresh[f]}, which makes the used \texttt{f} variable in
    the rules denote a fresh variable instance for each rewrite;
  \item the replacement (or \emph{contraction}) of the rules includes the fragment \texttt{\#Q[f]},
    which substitutes the variable matched in that position with the new fresh variable~\texttt{f}.
  \end{enumerate}

\end{itemize}
If we try to normalize the same term as before, CRSX outputs what is shown in
Figure~\ref{fig:N-out}.  Notice how the bound variables from the X program are now converted to
field tags, which are free variables in the CRSX representation of the nested-relational algebra.


\begin{figure*}[t]\small
\begin{verbatim}
// R scheme: basic traditional relational optimizations.
R[(
RemoveDepMap[Weak[#dop]] : Dep[id1.MapConcat[#dop[], id1]] → #dop ;
Productize[Weak[#op1]] : MapConcat[Dep[id.#op1[]], #op2] → Product[#op1, #op2] ;
)]
\end{verbatim}
\caption{\emph{R.crs}---simple relational optimizations.}\label{fig:R}
\vspace{1pc}
\footnotesize
\begin{verbatim}
Algebraic[
 (Dep id .
  Map[
   (Dep id1 . Call["plus", Concat[Extract[id1, v"$x"], Extract[id1, v"$y"]]]),
   Select[
    (Dep id1_1 . Call["eq", Concat[Extract[id1_1, v"$x"], Extract[id1_1, v"$y"]]]),
    Product[
     Map[(Dep id1_2 . Tuple[ACons[(v"$y" id1_2), ANil]]), Call["child", Call["doc", Empty]]],
     Product[
      Map[(Dep id1_3 . Tuple[ACons[(v"$x" id1_3), ANil]]), Call["child", Call["doc", Empty]]],
      id]]]])]
\end{verbatim}
\caption{Rewritten version of sample query.}\label{fig:R-sample}
\end{figure*}

\section{Rewriting}\label{R}

The purpose of using a relational algebra intermediate language is usually to rewrite queries to a
more optimized form.  Figure~\ref{fig:R} contains a few such standard optimizations:
\begin{itemize}

\item The rules are not related to a compilation scheme and can thus ``fire'' at any time.  This
  means that implementations should do some kind of completion procedure~\cite{KnuthBendix:1970} to
  ensure that the rules are applied properly, for example inserting a check for the application of
  these rules when a \verb|Dep| term in one of the involved constructors is created.

\item The \texttt{RemoveDepMap} rule includes the special \texttt{Weak[\#dop]} option.  This option
  states that the pattern for the \texttt{\#dop} meta-variable \emph{may} have an incomplete list of
  binders to indicate that the missing binders do not occur (free) in matching subterms.  We exploit
  this in the pattern by not listing the one bound variable, \texttt{id1}, as an argument to the
  meta-application of \texttt{\#dop} to ensure that the subterm matching the meta-application does
  not contain \texttt{id1}, which permits us to use it in the replacement without providing a
  substitution for \texttt{id1}.  Thus the rule states that nesting of a dependent operator can be
  ignored if the dependent operator does not in fact depend on the nested tuple.

\item Similarly, the \texttt{Productize} rule states that if the dependent operator of nesting is
  independent of the dependency then the two can be rewritten to a simple product.  The final
  rewrite here merely permits delaying tests, which allows combining the tests.

\end{itemize}
We shall not show any specific rules that perform annotation but just mention that they typically
take the form of an ``annotation scheme'' like
\begin{verbatim}
   {id:#cType}Type[id] → #cType
\end{verbatim}
where an environment in \verb|{}|s is used to pass the types of variables to the individual subterms
and construct their type (for the specifics of the CRSX environment notation see the appendix).  For
more complex analyses, inference rules like
\begin{displaymath}
  \dfrac
  {
    \rho\vdash p_2 : t_2
    \quad
    \rho+(i:t_2)\vdash p_1 : t
  }
  {\rho\vdash\mathtt{Map[Dep}~i\mathtt{.}p_1\mathtt{,}\,p_2\mathtt{]} : t}
\end{displaymath}
are encoded with generated rule schemes that rewrite terms like $\{\rho\}\texttt{⊢?}[p]$ to
$\texttt{⊢!}[t]$ when the rules can prove $\rho ⊢ p : t$, which is encoded for the above
rule as follows (shown without options):
\begin{verbatim}
   {#rho}"⊢?"[Map[Dep i.#p1[i], #p2]]
     → {#rho}"⊢??"[∀ i."⊢?1"[i, #p1[i], #p2, {#rho}"⊢?"[#p2]]] ;
   {#rho}"⊢?1"[i, #p1, #p2, "⊢!"[#t2]]
     → {#rho}"⊢?2"[i, #p1, #p2, #t2, {#rho;i:#t2}"⊢?"[#p1]] ;
   {#rho}"⊢?2"[i, #p1, #p2, #t2, "⊢!"[#t]] → "⊢!!"[i, #p1, #p2, #t2, #t] ;
   {#rho}"⊢??"[∀ i."⊢!!"[i, #p1[i], #p2, #t2, #t]] → "⊢!"[#t] ;
\end{verbatim}
that introduce helper translation schemes to build the proof of the inference rules in a strictly
deterministic left to right fashion.  (This is automated by a CRSX meta-rewrite system in the real
compiler.)


\begin{figure*}[p]\footnotesize
\begin{verbatim}
// E scheme: emit executable "pipeline code" from nested-relational algebra.
E[(

// Main program is a pipe from input cursor to output handler.
E[Algebraic[Dep id.#op[id]]] → TMain[in out.TPipe[h.TCopy[in, h], c.E2[#op[c], out]]] ;

// E2[ operator, handler ] generates code for operator to send the result value to the handler.

-[Free[h]]     : E2[Concat[#1, #2]   , h] → TSeq[E2[#1, h], E2[#2, h]] ;

-[Free[h]]     : E2[Empty            , h] → TNoop ;

-[Free[h]]     : E2[Literal[#N]      , h] → TLiteral[#N, h] ;

-[Free[h]]     : E2[Element[#1, #2]  , h]
 → TMakeElement[labelh.E2[#1, labelh], contenth.E2[#2, contenth], h] ;

-[Free[h]]     : E2[Call[#fun, #args], h] → TCall[#fun, argsh . E2[#args, argsh], h] ;

-[Free[c,f,h]] : E2[Extract[c,f]     , h] → TPick[c, f, h] ;
-[Free[h]]     : E2[Tuple[#fs]       , h] → MkT[#fs, TDNil, vh.TNoop, h] ;

// Helper to generate tuples.
-[Free[f,h]] : MkT[ACons[f #, #fs], #td, vh.#e[vh], h]
 → MkT[#fs, TDCons[f, #td], vh.TSeq[E2[#,vh], #e[vh]], h] ;
-[Free[h]]   : MkT[ANil           , #td, vh.#e[vh], h] → TMakeTuple[#td, vh.#e[vh], h] ;

-[Free[h]] : E2[Conditional[#,#1,#2], h]
 → TSwitch[caseh . E2[#, caseh], TCase[True, E2[#1, h], TOtherwise[E2[#2, h]]]] ;

// Basic queries.

-[Free[h]] : E2[Map[Dep id.#dop[id], #], h] → TPipe[h1.E2[#, h1], c1.E2[#dop[c1], h]] ;

-[Free[h]] : E2[Select[Dep id.#dop[id], #], h]
 → TPipe[h1.E2[#, h1],
  c1.TSwitch[caseh . E2[#dop[c1], caseh], TCase[True, TCopy[c1, h], TOtherwise[TEmpty]]]] ;

-[Free[h]] : E2[MapConcat[Dep id.#dop[id], #], h]
 → TPipe[h1.E2[#, h1], c1.TPipe[h2.E2[#dop[c1], h2], c2.TMerge[c1, c2, h]]] ;

-[Free[h]] : E2[Product[#1, #2], h]
 → TPipe[h2.TMakeElement[lh.TLiteral['Columns', lh], ch.E2[#2,ch],
  c2.TPipe[h1.E2[#1, h1], c1.TPipe[h2.TCall["child", nh.TCopy[c2,nh]], c3.TMerge[c1, c3, h]]]]] ;

)]
\end{verbatim}
  \caption{\emph{E.crs}---emit code.}\label{fig:E}
\end{figure*}

\section{Code Emission}\label{E}

The generated code will use data flow macros, as is established practice for such compilers, but
using higher-order terms.  The rules for code emission are shown in Figure~\ref{fig:E}, and
correspond closely to the usual operational semantics of the nested-relational operators:
\begin{itemize}

\item The top level emission translation scheme is \verb|E|, which creates a ``main'' target program
  with explicit binders for the input and output channels.

\item The body of the main program is a ``pipe,'' which connects the input to the program and the
  program to the output.  It is implemented by \texttt{TPipe}, the workhorse that creates a pair of
  a handler and a cursor, where the cursor is iterated over once for each value received by the
  handler: this iteration is what enables the identification of ``tuple'' with the usual ``frame''
  because a tuple of values sent to a handler is the same as the frame of registers received by the
  iteration code through the cursor.

\item The subsceheme \verb|E2| translates each algebraic construct to an explicit data flow.
  Concatenation, for example, is achieved by doing the code in sequence with output to the same
  handler.

\item Function call is interesting as the data flow architecture dictate that the way to instantiate
  a new frame for executing the function is to create a handler to send the function's arguments to
  and then invoke the function including the handler to which the result should be sent.

\item Records (in relational algebra called ``tuples'') are represented as terms by recursive lists
  with a member per field.

\item We use CRSX variables as ``data flow register'' represented by field tags, cursors
  representing the current value of an iteration, and handlers that can receive values for
  iteration; one can say that we use free CRSX variables similar to the way traditional code
  generation uses an ``infinite register model.''

\item Control instructions combine existing pieces of code; the \texttt{TSwitch} code generator is
  the only branch construct that receives a single value on a handler and delegates to the branch
  marked with that value (or, for elements, the tag of the value).

\item The data manipulation macros correspond to usual register lookup, frame copy, and frame merge
  operations.

\item The last rules show how relational algebraic operators are translated into pipes and merges.

\end{itemize}
Running our example through code emission
gives the result shown in Figure~\ref{fig:E-sample}.\footnote{The mechanisms used are rather crude.
  Notice for example how the \texttt{Product} operators result in the code building element
  containers to cache the columns.}

\begin{figure*}[t]\scriptsize
\begin{verbatim}
TMain[
 in out .
  TPipe[
   h. TCopy[in, h],
   id .
    TPipe[
     h1 .
      TPipe[
       h1_1 .
        TPipe[
         h2 .
          TMakeElement[
           lh. TLiteral[Columns, lh],
           ch .
            TPipe[
             h2_1. TMakeElement[lh_1. TLiteral[Columns, lh_1], ch_1. TCopy[id, ch_1]],
             c2 .
              TPipe[
               h1_2 .
                TPipe[
                 h1_3. TCall["child", argsh. TCall["doc", argsh_1. TNoop, argsh], h1_3],
                 id1. TMakeTuple[TDCons[v"$x", TDNil], vh. TSeq[TCopy[id1, vh], TNoop], h1_2]],
               c1. TPipe[h2_2. TCall["child", nh. TCopy[c2, nh]], c3. TMerge[c1, c3, ch]]]]],
         c2_1 .
          TPipe[
           h1_4 .
            TPipe[
             h1_5. TCall["child", argsh_2. TCall["doc", argsh_3. TNoop, argsh_2], h1_5],
             id1_1. TMakeTuple[TDCons[v"$y", TDNil], vh_1. TSeq[TCopy[id1_1, vh_1], TNoop], h1_4]],
           c1_1. TPipe[h2_3. TCall["child", nh_1. TCopy[c2_1, nh_1]], c3_1. TMerge[c1_1, c3_1, h1_1]]]],
       id1_2 .
        TSwitch[
         caseh .
          TCall["eq", argsh_4. TSeq[TPick[id1_2, v"$x", argsh_4], TPick[id1_2, v"$y", argsh_4]], caseh],
         TCase[True, TCopy[id1_2, h1], TOtherwise[TEmpty]]]],
     id1_3 .
      TCall["plus", argsh_5. TSeq[TPick[id1_3, v"$x", argsh_5], TPick[id1_3, v"$y", argsh_5]], out]]]]
\end{verbatim}
  \caption{Sample emitted code.}\label{fig:E-sample}
\end{figure*}

One important issue that we have to resolve in practice is to get all the optimizations to be
applied \emph{before} code generation.  This requires a study of the critical pairs of the system.
The system as presented here, for example, has an overlap between the \texttt{RemoveDepMap}
optimization rule and the \verb|E2| \verb|MapConcat| rule.  The solution in this case is \emph{not}
traditional completion as that will effectively mean that all optimizations have to be equivalently
implemented in the IL and TL but rather we simple block the cases for code generation that can be
handled by an optimization rule.  So the actual \verb|E2| \verb|MapConcat| rule looks like this:
\begin{verbatim}
-[Free[h]] : E2[MapConcat[Dep id.$[NotMatch,#dop[],#dop[id]], #], h]
→ TPipe[h1.E2[#, h1], c1.TPipe[h2.E2[#dop[c1], h2], c2.TMerge[c1, c2, h]]] ;
\end{verbatim}
(In practice, such choices are delegated to an analysis phase which drops cookies of some kind into
the term to serve as enablers of the overlapping rewrite steps.)


\section{Discussion}\label{Wrap}

At the end what remains is to put all the pieces together.  The driver is the top-level X symbol
introduced by parsing.  We add a small ``driver file'' that essentially rewrites
$\mathtt{E[N[}q\mathtt{]]}$ for queries $q$.

I have found that this kind of architecture is quite consistent with what compiler development teams
expect even if the notations used are of a more formal nature than most developers usually work
with.  The support for traditional ``compiler block diagrams'' like the one in the introduction,
where the fact that each analysis and translation is specified independently makes using a
structured approach realistic.  The chaotic nature of the resulting execution of the specification
comes out as an advantage and our implementation using a standard functional innermost-needed
strategy often ends up interleaving the stages of the compilation in interesting ways, for example
eliminating dead code before type checking, usually making mistakes in dependencies blatantly
obvious.  (Indeed, rewriting permits tweaking the reduction order or using tricks such as completion
to discover bad dependencies early.)  However, debugging of rule systems is very different from
usual debugging in that mistakes show up as ``unsimplified blobs or term,'' which is different from
actual crashes (and requires strict discipline in naming the various modular components in a
globally identifyable way, something we have side-stepped in this brief presentation).

Although we have not covered it here, we have observed that the rewrite systems obtained can even
themselves be translated mechanically to low-level code, making it feasible to implement the actual
production compiler direclty from the rewrite rules.  Important factors in this has been the disciplined use of systems
that can be transformed into orthogonal constructor systems, for which a table-driven normalizing
strategy can be used in almost all cases (there is a performance penalty for some substitution
cases).

The CRSX system implements higher order rewriting fully in the form of CRS, thus can handle full
substitution and thus express transformations such as inlining.  However, it turns out that many
specific systems share with the small ones presented here the property that they use only ``explicit
substitution'' style rewrites, which only permits observing variables~\cite{BlooRose:rta1996}.
Indeed it seems that the fact that the approach is \emph{not} functional or a full logical framework
is an advantage: the expressive power of explicit substitution is strictly smaller (in a complexity
sense) than general functions.

Finally, a crucial component in using rewriting for specifying large rule sets as is the case in the
real compiler is the strict shape requirements on rules: basically every aspect of a rule that is
not strictly linear and only substitutes bound variables for bound variables without any constraints
is an error unless it is explicitly requested: this purely syntactic approach catches numerous
errors early.

\paragraph*{Related Work.}

The area of verifying a compiler specification is well established using both hand-written and
mechanical proofs~\cite{Dave:sen2003}.  Work has also been done on linking correct compiler
specification and implementations using generic proof theoretic
tools~\cite{OkumaMinamide:aplas2003}.  Tools supporting mechanical generation of compilers from
specifications, such as SDF+ASF~\cite{Brand+:toplas2002} and Stratego~\cite{Bravenboer+:fi2006},
have focused on compilers restricted to first-order representations of intermediate languages used
by the compiler and on using explicit rewriting strategies to guide compilation.  Our goal is the
opposite: to only specify dependencies between components of the compiler and leave the actual
rewriting strategy to the system (in practice using analysis-guided rule transformations coupled
with a generic normalizing strategy).

We are only aware of one published work that uses higher order features with compiler construction,
namely the work by Hickey and Nogin on specifying compilers using logical
frameworks~\cite{HickeyNogin:hosc2006}.  The resulting specification looks very similar to ours, and
indeed one can see the code synthesis that could be done for their logic system as similar to the
code generation we are employing.  Also, both systems employ embedded source language syntax and
higher-order abstract syntax.  However, there are differences as well.  First, CRSX is explicitly
designed to implement just the kind of rewrite systems that we have described, and is tuned to
generate code that drives transformation through lookup tables.  Second, variables are first class
in CRSX and not linked to meta-level abstraction, thus closer to the approach used by explicit
substitution for CRS~\cite{BlooRose:rta1996} and ``nominal''
rewriting~\cite{FernandezGabbay:ic2007}.  This permits us, for example, to use an assembly language
with mutable registers.  Third, we find that the focus on local rewriting rules is easier to explain
to compiler writers, and the inclusion of environments and inference rules in the basic notation
further helps.  Finally, the CRSX engine has no assumed strategy so we find the notion of local
correctness easier to grasp.

\paragraph*{What's Next?}

With CRSX we continue to experiment with pushing the envelope for supporting more higher-order
features without sacrificing efficiency.

An important direction is to connect with nominal rewriting and understand the relationship between
what the two formalisms can express.

Another interesting direction for both performance and analysis is to introduce explicit
\emph{weakening} operators that ``unbind'' a given bound variable in a part of its scope.  While
used in this way with explicit substitution~\cite{RoseBlooLang:ibm2009,KesnerRenaud:mfcs2009}, the
interaction with higher-order rewriting is not yet clear.

In companion papers we explain the details of the translation from the supported three forms of
rules, ``recursive compilation scheme,'' ``chaotic annotation rules,'' and ``deterministic inference
rules,'' into effective native executables, and we explain annotations that make it feasible to avoid
rewriting-specific static mistakes.

\paragraph*{Acknowledgements.}

The author is grateful for insightful comments by the anonymous referees including being made aware
of the work in logical frameworks.


\bibliography{crs}

\end{document}